\documentclass[twocolumn]{29onr_art}
\usepackage{ulem}
\usepackage[dcucite]{harvard}
\usepackage{graphicx}
\usepackage{fixltx2e}
\usepackage[intlimits]{amsmath}
\usepackage{color}
\usepackage[colorlinks=true,
            urlcolor=blue,
            linkcolor=docgreen,
            citecolor=docgreen,
            bookmarks=true,
            pdftitle={Numerical Simulation of Internal Tide Generation at a Continental Shelf Break},
            pdfauthor={L.K. Brandt},
            pdfsubject={Proceedings of the 29th Symposium on Naval Hydrodynamics},
            pdfkeywords={Internal Waves, Solitons, Beams, Numerical Simulations, High Performance Computing},
            bookmarksopen=false,
            pdfpagemode=UseNone]{hyperref}

\definecolor{docgreen}{rgb}{0,.8,0}

\newcommand\Rey{\mbox{\textit{Re}}}  

\def\be{\begin{eqnarray}}
\def\ee{\end{eqnarray}}
\def\benl{\begin{eqnarray*}}
\def\eenl{\end{eqnarray*}}

\newcommand{\nwc}{\newcommand}
\nwc{\bm}{\boldmath}
\nwc{\m}{\mbox}
\nwc{\ubm}{\unboldmath}
\nwc{\bmU}{\m{\bm$U$\ubm}}
\nwc{\bmX}{\m{\bm$X$\ubm}}
\nwc{\bmu}{\m{\bm$u$\ubm}}
\nwc{\bmx}{\m{\bm$x$\ubm}}
\nwc{\bmz}{\m{\bm$z$\ubm}}
\nwc{\bmv}{\m{\bm$v$\ubm}}
\nwc{\bmw}{\m{\bm$w$\ubm}}
\nwc{\bmW}{\m{\bm$W$\ubm}}
\nwc{\bmn}{\m{\bm$n$\ubm}}
\nwc{\bmG}{\m{\bm$G$\ubm}}
\nwc{\bmF}{\m{\bm$F$\ubm}}
\nwc{\bmI}{\m{\bm$I$\ubm}}
\nwc{\bmN}{\m{\bm$N$\ubm}}
\nwc{\bmP}{\m{\bm$P$\ubm}}
\nwc{\bmcalP}{\m{\bm $\cal P$\ubm}}
\nwc{\bmV}{\m{\bm$V$\ubm}}
\nwc{\bmS}{\m{\bm$S$\ubm}}

\begin{document}
\pagenumbering{arabic}

%
%
%
%
%
%
%
%
\title{Numerical Simulation of Internal Tide Generation at a Continental Shelf Break}
%
%
%
\author{Laura K. Brandt, James W. Rottman, Kyle A. Brucker and Douglas G.Dommermuth}
\affiliation{Naval Hydrodynamics Division, Science Applications International Corporation
10260 Campus Point Drive, San Diego, CA  92121, USA}
\maketitle
%
%
\section{ABSTRACT}

A fully nonlinear, three-dimensional numerical model is developed for the simulation of tidal flow over arbitrary bottom topography in an ocean with realistic stratification. The model is capable of simulating accurately the generation of fine-scale internal wave tidal beams, their interaction with an ocean thermocline and the subsequent generation of solitary internal waves that propagate on this thermocline. Several preliminary simulation results are shown for uniform and non-uniform flow over an idealized two-dimensional ridge, which are compared with linear theory, and for flow over an idealized two-dimensional continental shelf.

\section{INTRODUCTION} \label{sec:introduction}

It is important to have a good understanding of the ocean environment in which surface and subsurface ships operate.  In particular, submarines operating in the littoral ocean environment can be significantly affected by the presence of large-amplitude internal waves.  A generation mechanism for these waves is the motion of the barotropic tide over continental shelf breaks, as discussed, for example, by \citeasnoun{pingree1989}, \citeasnoun{holloway2001}, \citeasnoun{Lien2001}, and \citeasnoun{garrett2007}.  The ultimate objective is to produce a forecast model for the generation and propagation of large amplitude internal waves in a realistic ocean in the regions about a continental shelf.

The numerical modeling of this generation process is difficult because of the complexity of the topography, the complicated structure of the ocean stratification and currents, and the wide range of spatial and temporal scales.  When the continental slope is near a critical value, which occurs often, a fine-scale internal wave beam is generated that must be accurately resolved in a domain that has the very large length scale of the ocean shelf region.  Also, it is possible for this internal wave beam to produce internal solitary waves by the interaction of the beam with a moderately strong ocean thermocline, as described by \citeasnoun{gerkema2001} and \citeasnoun{mauge2008}.

In this paper, we develop a numerical scheme capable of accurately simulating the generation of these beams with realistic ocean stratification and bottom topography. The numerical method is fully nonlinear, and uses a technique similar to the Cartesian-grid free-surface capturing code called Numerical Flow Analysis (NFA) of \citeasnoun{DOWRWHYSAV07} and \citeasnoun{rottman2010}, but modified to handle nonlinear background stratification and ocean bottom topography. The scheme is sufficiently robust to simulate the generation of the internal wave beams, the interaction of these internal wave beams with an ocean thermocline, and the subsequent propagation of the generated internal solitary waves shoreward over realistic bottom topography. 
 
This paper also describes some preliminary simulations for stratified flow over idealized two-dimensional ridges, which are compared with linear theory, and a two-dimensional shelf break, including the generation of internal wave beams and the interaction of these beams with idealized ocean thermoclines.

\section{THE NUMERICAL MODEL} \label{sec:numericalmodel}

We consider nonlinear, three-dimensional, stratified fluid flow over bottom topography. A Cartesian coordinate system ($x$, $y$, $z$) is used with $z$ as the vertical coordinate and ($x$, $y$) as the horizontal coordinates. The background stratification is assumed variable in the vertical but homogeneous in the horizontal. Typically, we will impose a background stratification that represents an ocean with a seasonal thermocline. The background current is assumed to be forced by a barotropic tide. 

The computer code Numerical Flow Analysis (NFA), \citeasnoun{DOWRWHYSAV07}, 
originally designed to provide turnkey capabilities to simulate the free-surface flow around ships, has been extended to have the ability to perform high-fidelity stratified sub-surface calculations. The governing equations are formulated on a Cartesian grid thereby eliminating complications associated with body-fitted grids. The sole geometric input into NFA is a surface panelization of the ship and/or bottom. No additional gridding beyond what is used already in potential-flow methods and hydrostatics calculations is required. The ease of input in combination with a flow solver that is implemented using parallel-computing methods permit the rapid turnaround of numerical simulations of high-\Rey \hspace{1pt} stratified fluid interactions with a complex bottom.

The grid is stretched along the Cartesian axes using one-dimensional elliptic equations to improve resolution near the bottom and the mixing layer. Away from the bottom and the mixing layer, where the flow is less complicated, the mesh is coarser. Details of the grid-stretching algorithm, which uses weight functions that are specified in physical space, are provided in \citeasnoun{KnuppSteinberg1993}.

\subsection{Governing Equations} 

Consider a turbulent flow in a stratified fluid.  Physical quantities are normalized by characteristic velocity ($U_0$), length ($L_0$),  time ($L_0/U_0$), density ($\rho_0$), and pressure ($\rho_0 U_0^2$) scales.  Let $\rho$ and $u_i$, respectively denote the normalized density and three-dimensional velocity field as a function of normalized space ($x_i$) and normalized time ($t$).  
The conservation of mass is
\begin{equation} 
\label{eq:mass1} 
\frac{\partial \rho}{\partial t} +\frac{\partial u_j \rho}{\partial x_j} = 0 \;\; . 
\end{equation}
For incompressible flow with no diffusion,
\begin{equation} 
\label{eq:density1} 
\frac{\partial \rho}{\partial t} +u_j \frac{\partial \rho}{\partial x_j} = 0 \;\; .  
\end{equation}
Subtracting \eqref{eq:density1} from \eqref{eq:mass1} gives a solenoidal condition for the velocity:
\begin{equation} 
\label{eq:solenoidal} 
\frac{\partial u_i}{\partial x_i} = 0 \;\; .
\end{equation}

The normalized density is decomposed in terms of the constant reference density plus small departures that are further split into a known mean perturbation ($\overline{\rho}_C$), a continuous fluctuation ($\tilde{\rho}_C$) due to the mean density gradient, and a fluctuation with a discontinuous jump in the density ($\tilde{\rho}_J$) corresponding to the bottom of the mixing layer:
\begin{eqnarray}
\label{eq:density2}
\rho = 1  + \gamma_C \overline{\rho}_C\! \left(x_3\right) + \gamma_C \tilde{\rho}_C\!\left(x_i,t\right)+ \gamma_J \tilde{\rho}_J\!\left(x_i,t\right)  \; .
\end{eqnarray}
\noindent $\gamma_C$ and $\gamma_J$ quantify the magnitudes of the density fluctuations for the continuous and discontinuous portions, respectively : 
\begin{eqnarray}
\gamma_C & = & \frac{L_0}{\rho_0} \left. \frac{\partial \overline{\rho}}{\partial x_3} \right|_0 \\
\label{eq:density3}
\gamma_J & = &  \frac{\Delta \rho}{\rho_0}  \; .
\label{eq:density4}
\end{eqnarray}
\noindent Here, $(\partial \overline{\rho}/\partial x_3)_0$ is the dimensional characteristic mean-density gradient and $\Delta \rho$ is the dimensional density jump.   The density fluctuations are split into two parts because they require different theoretical and numerical treatments.

The splitting requires an additional equation that we choose as follows.
\begin{equation}
\label{eq:mass2} 
\frac{\partial \tilde{\rho}_J}{\partial t} +\frac{\partial u_j \tilde{\rho}_J}{\partial x_j} = 0 \;\; . 
\end{equation}
Substituting \eqref{eq:density2} and \eqref{eq:mass2} into \eqref{eq:mass1} gives
\begin{equation} 
\label{eq:mass3} 
\frac{\partial  \tilde{\rho}_C}{\partial t} +\frac{\partial u_j (\overline{\rho}_C+ \tilde{\rho}_C)}{\partial x_j} = 0 \;\; . 
\end{equation}

For an infinite Reynolds number, viscous stresses are negligible, and the conservation of momentum is
\begin{eqnarray} 
\label{eq:momentum1} 
\frac{ \partial \rho u_i}{\partial t}+\frac{\partial}{\partial x_j}
\left(\rho u_j u_i \right)  = -\frac{\partial p}{\partial x_i} - \frac{\delta_{i3}}{F_r^2} - \tau_i  \; , 
\end{eqnarray}
where $p$ is the normalized pressure and $\tau_i$ is a normalized stress that will act tangential to the surface of the bottom.  $\delta_{ij}$
is the Kronecker delta function.   $F_r$ is a Froude number:
\begin{equation}
\label{eq:RiB}
F_r \equiv \frac{g L_0}{U_0^2} \, ,
\end{equation}
\noindent where $g$ is the acceleration of gravity.  The Froude number is the ratio of inertial to gravitational forces.  As \citeasnoun{dommermuth08} discuss, the sub-grid scale stresses are modeled implicitly in \ref{eq:momentum1}.

The pressure, $p$ is then decomposed into the dynamic, $p_d$, and hydrostatic, $p_h$, components as
\begin{equation}
\label{eq:pressure}
p= p_d + p_h \; .
\end{equation}
The hydrostatic pressure is defined in terms of the reference density and the density stratification as follows.
\begin{equation}
\label{eq:hydrostatic}
\frac{\partial p_h}{\partial x_i} = -( 1 + \overline{\rho}_C) \frac{\delta_{i3}}{F_r^2}  \; .
\end{equation}

The substitution of \eqref{eq:density2} and \eqref{eq:hydrostatic} into \eqref{eq:momentum1} and using \eqref{eq:mass2} and \eqref{eq:mass3}  to simplify terms gives a new expression for the conservation of momentum:  
\begin{multline} 
\label{eq:momentum2} 
\frac{ \partial u_i}{\partial t}+\frac{\partial}{\partial x_j} \left( u_j u_i \right)  = -\frac{1} {1+ \gamma_C (\overline{\rho}_C+\tilde{\rho}_C)+\gamma_J \tilde{\rho}_J }  \\
\left[ \frac{\partial p_d}{\partial x_i}  + (\gamma_C \tilde{\rho}_C+\gamma_J \tilde{\rho}_J)   \, \frac{\delta_{i3}}{F_r^2} +\tau_i \right].
\end{multline}
If $\gamma_C << 1$ and $\gamma_J << 1$, a Boussinesq  approximation may be employed in the preceding equation to yield
\begin{multline} 
\label{eq:momentum3} 
\frac{ \partial u_i}{\partial t}+\frac{\partial}{\partial x_j} \left( u_j u_i \right)  = \\
-\left[ \frac{\partial p_d}{\partial x_i}  + (Ri_{B_C} \tilde{\rho}_C+ Ri_{B_J}  \tilde{\rho}_J)   \, \delta_{i3} +\tau_i \right] .
\end{multline}
\noindent where $Ri_{B_C}$ and $Ri_{B_J}$ are bulk Richardson numbers defined as
\begin{eqnarray}
Ri_{B_C} & \equiv & \frac{\gamma_C}{F_r^2} \tilde{\rho}_C = \frac{L_0}{\rho_0} \left. \frac{\partial \overline{\rho}}{\partial x_3} \right|_0 \frac{g L_0}{U_0^2} \\
\label{eq:RiBC}
Ri_{B_J} & \equiv & \frac{\gamma_J}{F_r^2}  \tilde{\rho}_J = \frac{\Delta \rho}{\rho_0}\frac{g L_0}{U_0^2} \, ,
\label{eq:RiBJ}
\end{eqnarray}
The bulk Richardson numbers are the ratios of buoyant to inertial forces for continuous and discontinuous density fluctuations. 

The momentum equations using either \eqref{eq:momentum2} or \eqref{eq:momentum3}  and the mass conservation equations \eqref{eq:mass2} and \eqref{eq:mass3} are integrated with respect to time.    The divergence of the momentum equations in combination with the solenoidal condition \eqref{eq:solenoidal} provides a Poisson equation for the dynamic pressure.  The dynamic pressure is used to project the velocity onto a solenoidal field and to impose a no-flux condition on the surface of the body.   The details of the time integration, the pressure projection, the formulations of the body boundary conditions, and the formulations of the inflow and outflow boundary conditions are described in the next three sections.

\subsection{Time Integration}

A second-order Runge-Kutta scheme is used to integrate with respect to time the field equations for the velocity and density.  During the first stage of the Runge-Kutta algorithm, a Poisson equation for the pressure is solved:
\begin{multline}
\label{eqn:poisson1}
\frac{\partial}{\partial x_i} \left[ \frac{1} {1+ \gamma_C (\overline{\rho}_C+\tilde{\rho}^k_C)+\gamma_J \tilde{\rho}^k_J } \right] \frac{\partial p_d^k}{\partial x_i} = \\\frac{\partial}{\partial x_i} \left(
\frac{u^k_i}{\Delta t}+R^k_i \right) \; ,
\end{multline}
where $R^k_i$ denotes the nonlinear convective, buoyancy, and stress terms in the momentum equation, \eqref{eq:momentum2}, at time step $k$.
$u^k_i$, $\tilde{\rho}^k_C$, $\tilde{\rho}^k_J$, and $p_d^k$ are, respectively, the velocity components, continuous fluctuating density, discontinuous fluctuating density, and dynamic pressure at time step $k$.  $\Delta t$ is the time step.
For the next step, this pressure is used to project the velocity onto a solenoidal field. The first prediction for the velocity field ($u^*_i$) is
\begin{multline}
\label{eqn:runge1}
u^*_i=u^k_i+ \Delta t \bigg( R^k_i \\
-\left. \left[ \frac{1} {1+ \gamma_C (\overline{\rho}_C+\tilde{\rho}^k_C)+\gamma_J \tilde{\rho}^k_J }   \right] \frac{\partial p_d^k}{\partial x_i} \right) \; .
\end{multline}
The densities are advanced using the mass conservation equations \eqref{eq:mass2} and  \eqref{eq:mass3}:
\begin{eqnarray}
\tilde{\rho}^*_C & = & \tilde{\rho}^{k}_C- \Delta t \frac{\partial} {\partial
x_j} \left( u^k_j (\overline{\rho}_C+\tilde{\rho}^k_C) \right) \\
\tilde{\rho}^*_J & = & \tilde{\rho}^{k}_J- {\rm VOF} \left( u^k_j,  \tilde{\rho}^k_J , \Delta t \right).
\end{eqnarray}
The advective terms for $\tilde{\rho}_C$ are calculated using a third-order finite-volume approximation, whereas the advection of $\tilde{\rho}_J$ is calculated using the Volume of Fluid (VOF) method.  A Poisson equation for the pressure is solved again during the second stage of the Runge-Kutta algorithm:
\begin{multline}
\label{eqn:poisson2}
\frac{\partial}{\partial x_i} \left[ \frac{1} {1+ \gamma_C (\overline{\rho}_C+\tilde{\rho}^*_C)+\gamma_J \tilde{\rho}^*_J } \right] \frac{\partial p_d^*}{\partial x_i} = \\\frac{\partial}{\partial x_i} \left(
\frac{u_i^*+u^k_i}{\Delta t}+R^*_i \right) \; ,
\end{multline}
$u_i$ is advanced to the next step to complete one cycle of the Runge-Kutta algorithm:
\begin{multline}
\label{eqn:runge2}
u^{k+1}_i=\frac{1}{2} \left( u^*_i + u^k_i \right)  +\frac{\Delta t}{2} \bigg( R^*_i  \\
 \left. -\left[ \frac{1} {1+ \gamma_C (\overline{\rho}_C+\tilde{\rho}^*_C)+\gamma_J \tilde{\rho}^*_J }  \right] \frac{\partial p_d^*}{\partial x_i}  \right) \;\; ,
\end{multline}
and the densities are advanced to complete the algorithm:
\begin{eqnarray}
\tilde{\rho}^{k+1}_C & = & \frac{\tilde{\rho^*}_C+\tilde{\rho}^{k}_C}{2} -
\frac{\Delta t}{2} \frac{\partial} {\partial x_j} \left( u^*_j
(\overline{\rho}_C+\tilde{\rho}^*_C) \right)   \nonumber \\
\tilde{\rho}^{k+1}_J & = & \tilde{\rho}^{k}_J- {\rm VOF} \left( \frac{u^*_j+u^k_j}{2},  \tilde{\rho}^*_J , \Delta t \right) \; .
\end{eqnarray}

\subsection{Enforcement of No-Flux Boundary Conditions}
 
  A no-flux condition is satisfied on the surface of the bottom using a finite-volume technique. 
\begin{equation}
\label{noflux}
u_i n_i = v_i n_i \; ,
\end{equation}
where $n_i$ denotes the unit normal to the body that points into the fluid and $v_i$ is the velocity of the bottom. If the bottom is not moving, $v_i=0$.  Cells near the bottom may have an irregular shape, depending on how the surface of the bottom cuts the cell.   Let $S_b$ denote the portion of the cell whose surface is on the bottom, and let $S_0$ denote the other bounding surfaces of the cell that are not on the bottom.  

Gauss's theorem is applied to the volume integral of \eqref{eqn:poisson1}:
\begin{multline}
\label{eqn:integral1}
\lefteqn{ \int_{S_0+S_b} ds   \left[ \frac{n_i} {1+ \gamma_C (\overline{\rho}_C+\tilde{\rho}^k_C)+\gamma_J \tilde{\rho}^k_J } \right]  \frac{\partial p_d^k}{\partial x_i}  =} \\
 \int _{S_0+S_b} ds \left( \frac{u^k_i n_i}{\Delta t}+R_i n_i \right)  .
\end{multline}
Here, $n_i$ denotes the components of the unit normal on the surfaces
that bound the cell.   Based on \eqref{eqn:runge1}, a Neumann condition is derived for the pressure on $S_b$ as follows 
\begin{multline}
\label{eqn:bc1}
\left[ \frac{n_i}{1+ \gamma_C (\overline{\rho}_C+\tilde{\rho}^*_C)+\gamma_J \tilde{\rho}^k_J} \right]
\frac{\partial p_d^k}{\partial x_i} = \\
-\frac{u^*_i n_i}{\Delta t} +\frac{u^k_i n_i}{\Delta t}+R_i n_i \;\; .
\end{multline}
The Neumann condition for the velocity \eqref{noflux} is substituted into the preceding equation to complete the Neumann condition for the pressure on $S_b$:
\begin{multline}
\label{eqn:bc2}
\left[ \frac{n_i}{1+ \gamma_C (\overline{\rho}_C+\tilde{\rho}^*_C)+\gamma_J \tilde{\rho}^k_J} \right]
\frac{\partial p_d^k}{\partial x_i} =\\
-\frac{v^*_i n_i}{\Delta t} +\frac{u^k_i n_i}{\Delta t}+R_i n_i \;\; .
\end{multline}
This Neumann condition for the pressure is substituted into the integral formulation in \ref{eqn:integral1}:
\begin{eqnarray}
\label{eqn:integral2}
\lefteqn{ \int_{S_0} ds   \left[ \frac{n_i} {1+ \gamma_C (\overline{\rho}_C+\tilde{\rho}^*_C)+\gamma_J \tilde{\rho}^k_J} \right]  \frac{\partial p_d^k}{\partial x_i}  =} \nonumber \\
& &  \int _{S_0} ds \left( \frac{u^k_i n_i}{\Delta t}+R_i n_i \right) +  \int _{S_b} ds  \frac{v^*_i n_i}{\Delta t}  \; .
\end{eqnarray}
This equation is solved using the method of fractional areas.  Details associated with the calculation of the area fractions are provided in \citeasnoun{sussman01} along with additional references.  Cells with a cut volume of less than 2\% of the full volume of the cell are merged with neighbors.  The merging occurs along the direction of the normal to the body. This improves the conditioning of the Poisson equation for the pressure.   As a result, the stability of the  projection operator for the velocity is also improved (see Equations \eqref{eqn:runge1} and \eqref{eqn:runge2}). 

\begin{figure*}[h!]
  \centerline{
     \includegraphics[width=0.35\linewidth]{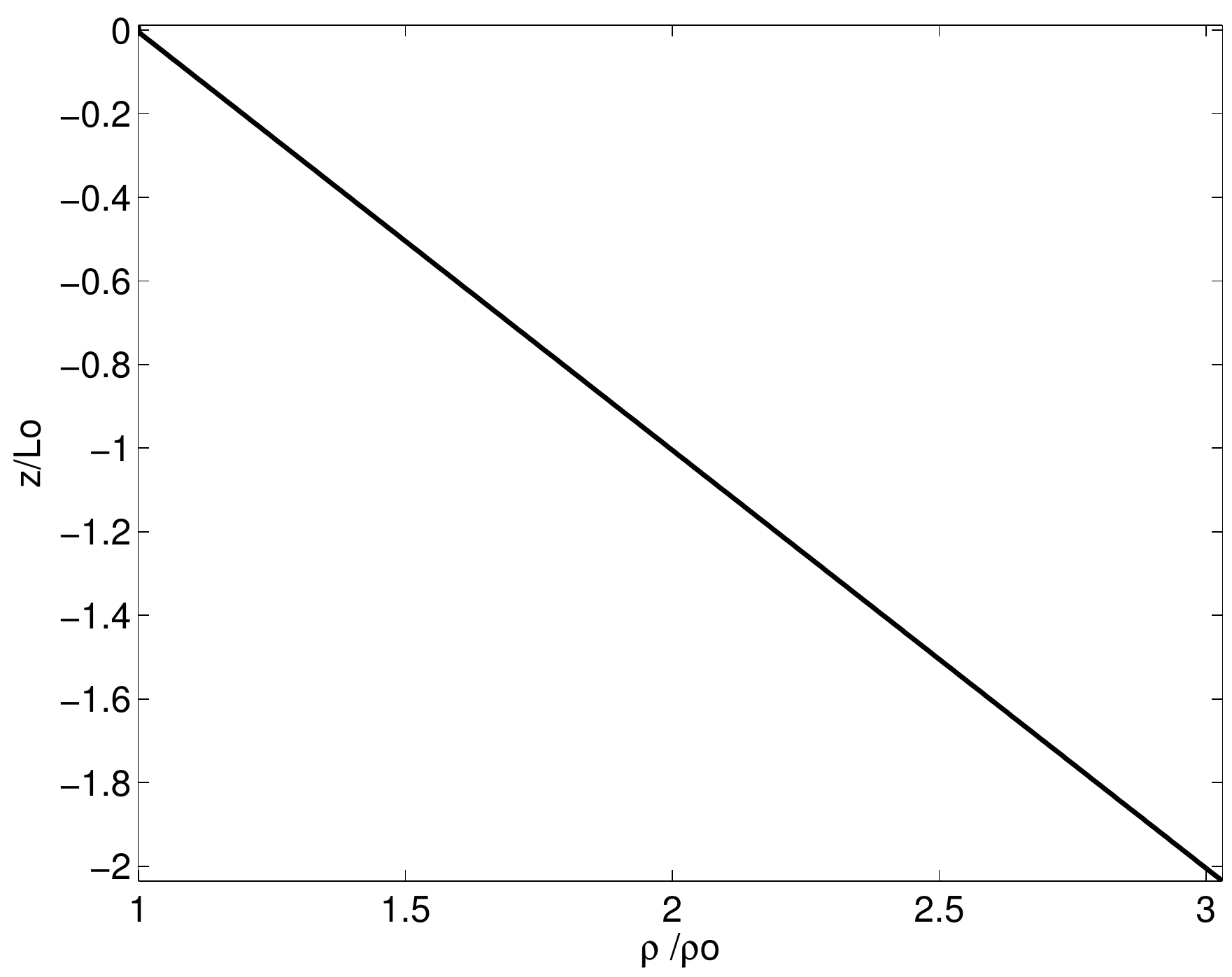}
     \hspace{0.5in}
     \includegraphics[width=0.35\linewidth]{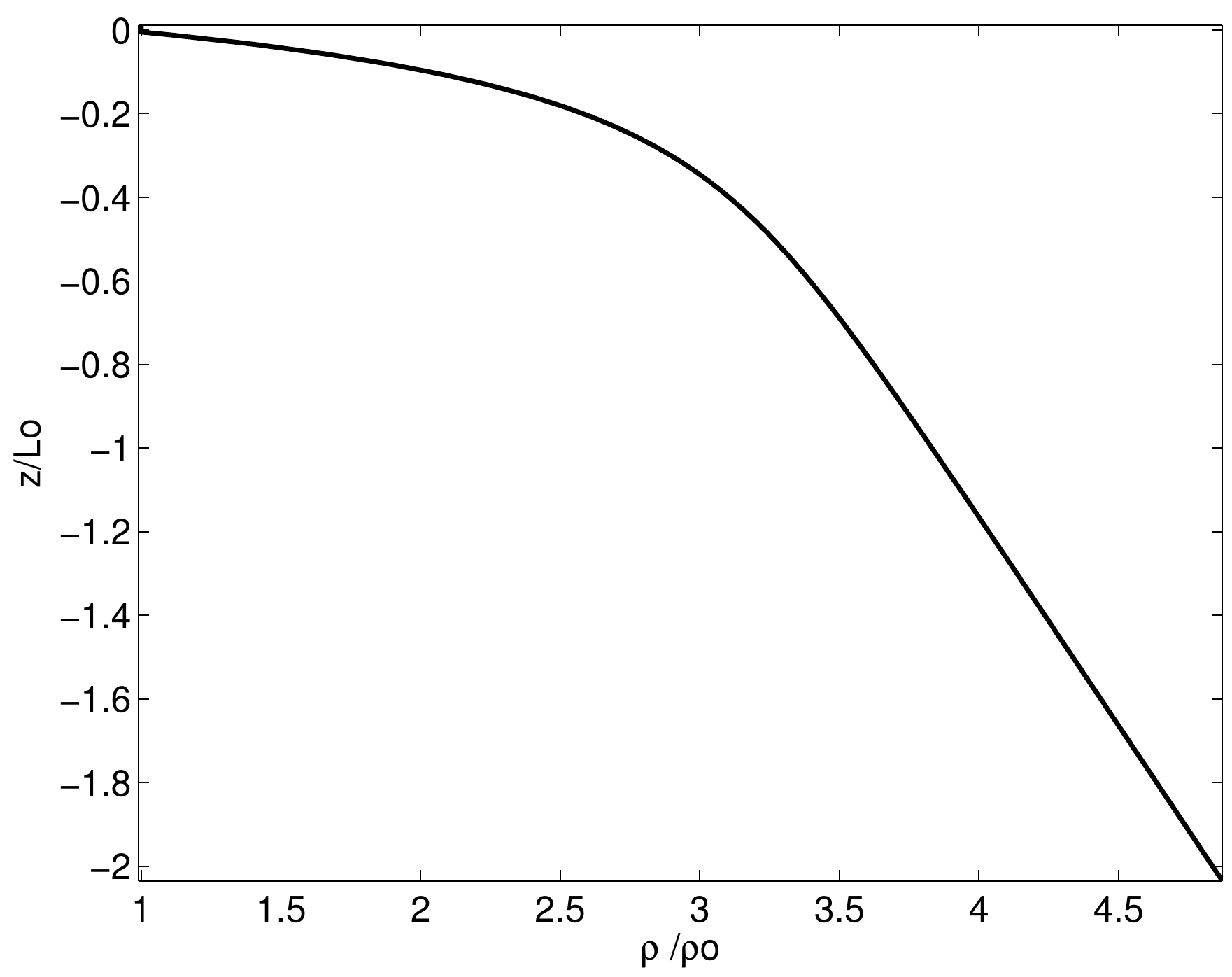}
   }
  \centerline{
  {\small (a)}\hspace{2.5in}{\small (b)}
  }
  \caption{Normalized density $\rho/\rho_0$ as a function of normalized depth $z/L_0$, where $L_0 = H - h_0$: (a) uniform stratification; and (b) nonuniform stratification representing a typical seasonal thermocline.}
\label{fig:Density_hill}
\end{figure*}

\begin{figure}[h!]
     \centerline{(a)\includegraphics[width=0.85\linewidth]{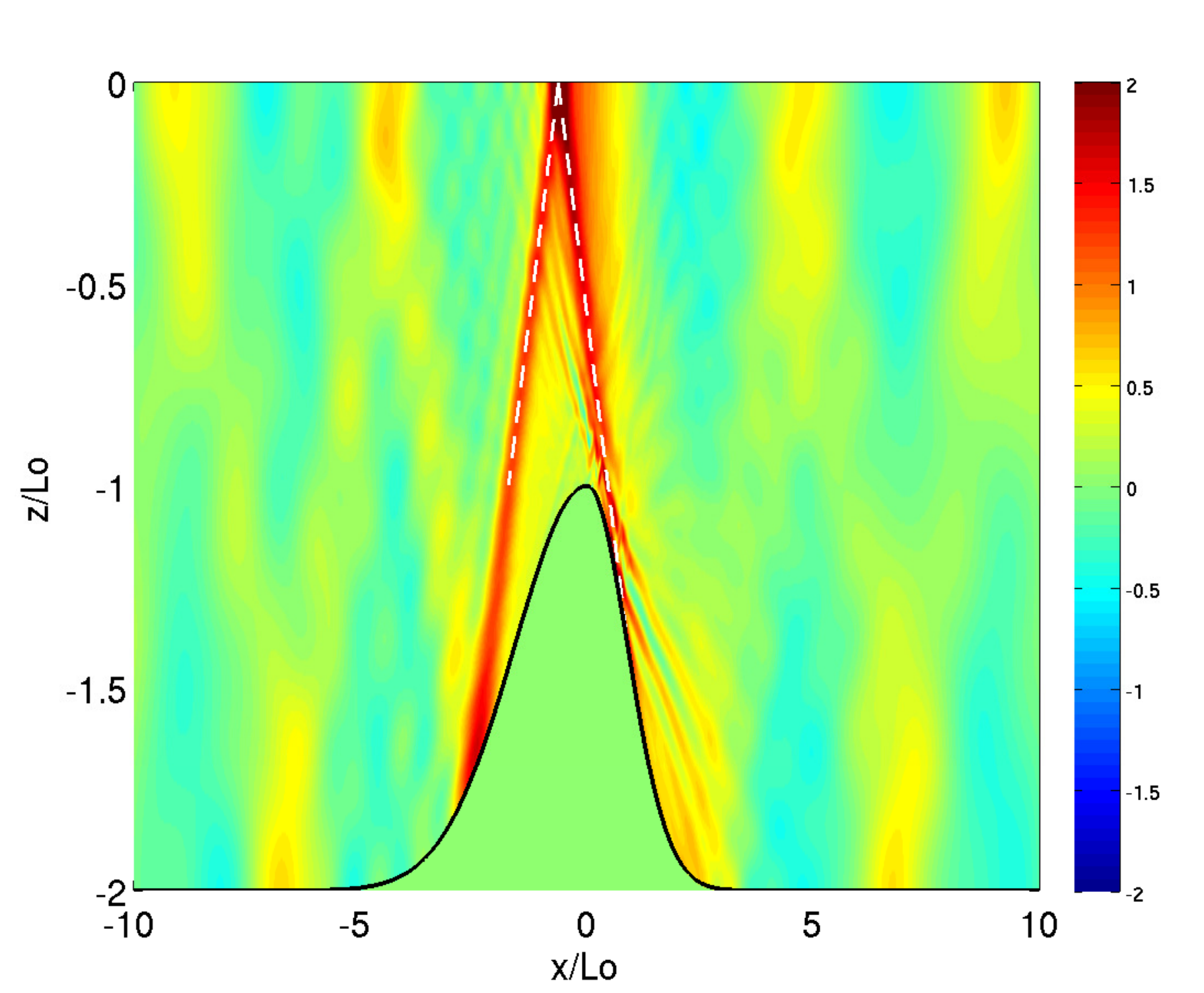}}
     \centerline{(b)\includegraphics[width=0.85\linewidth]{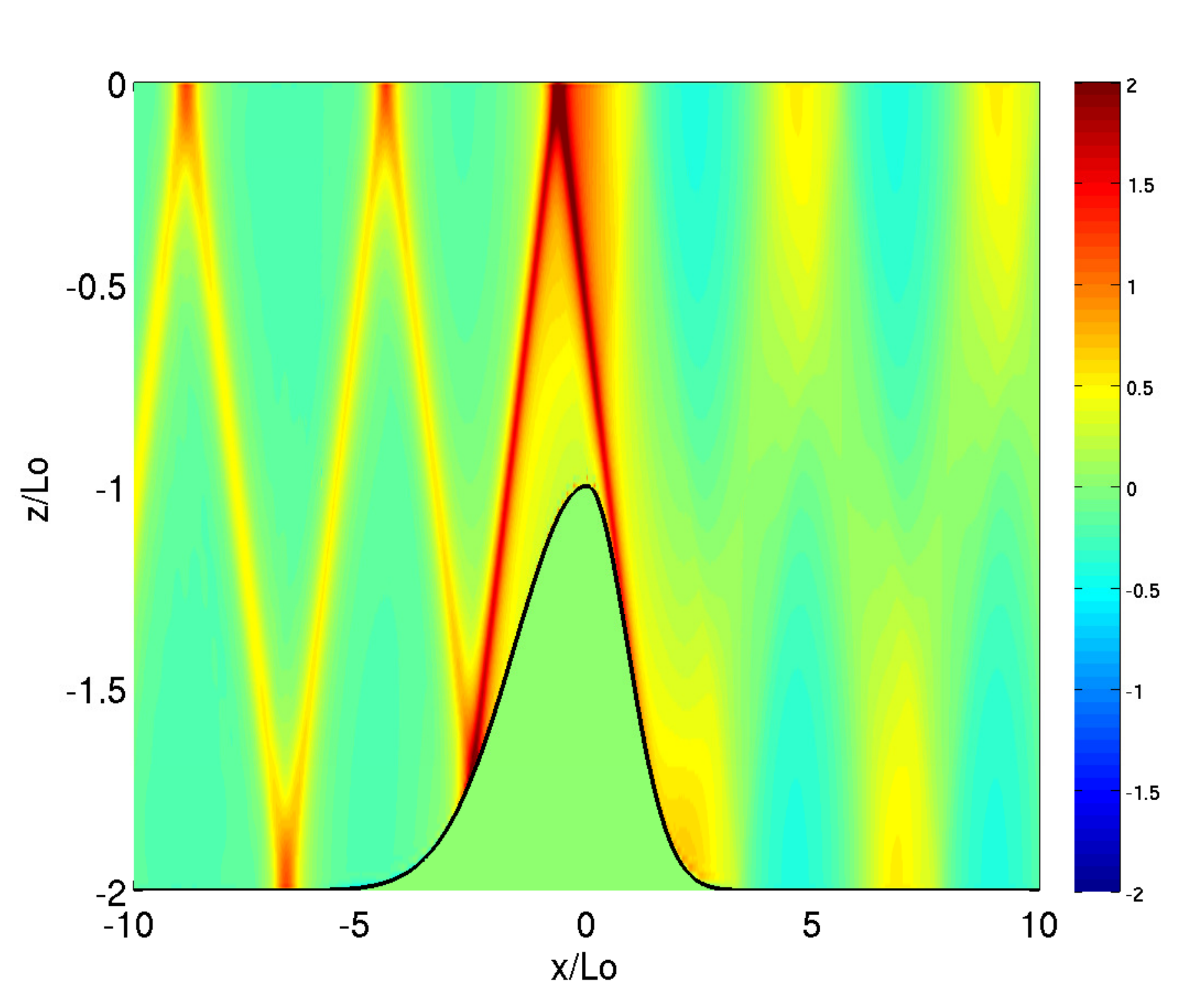}}
     \centerline{(c)\includegraphics[width=0.85\linewidth]{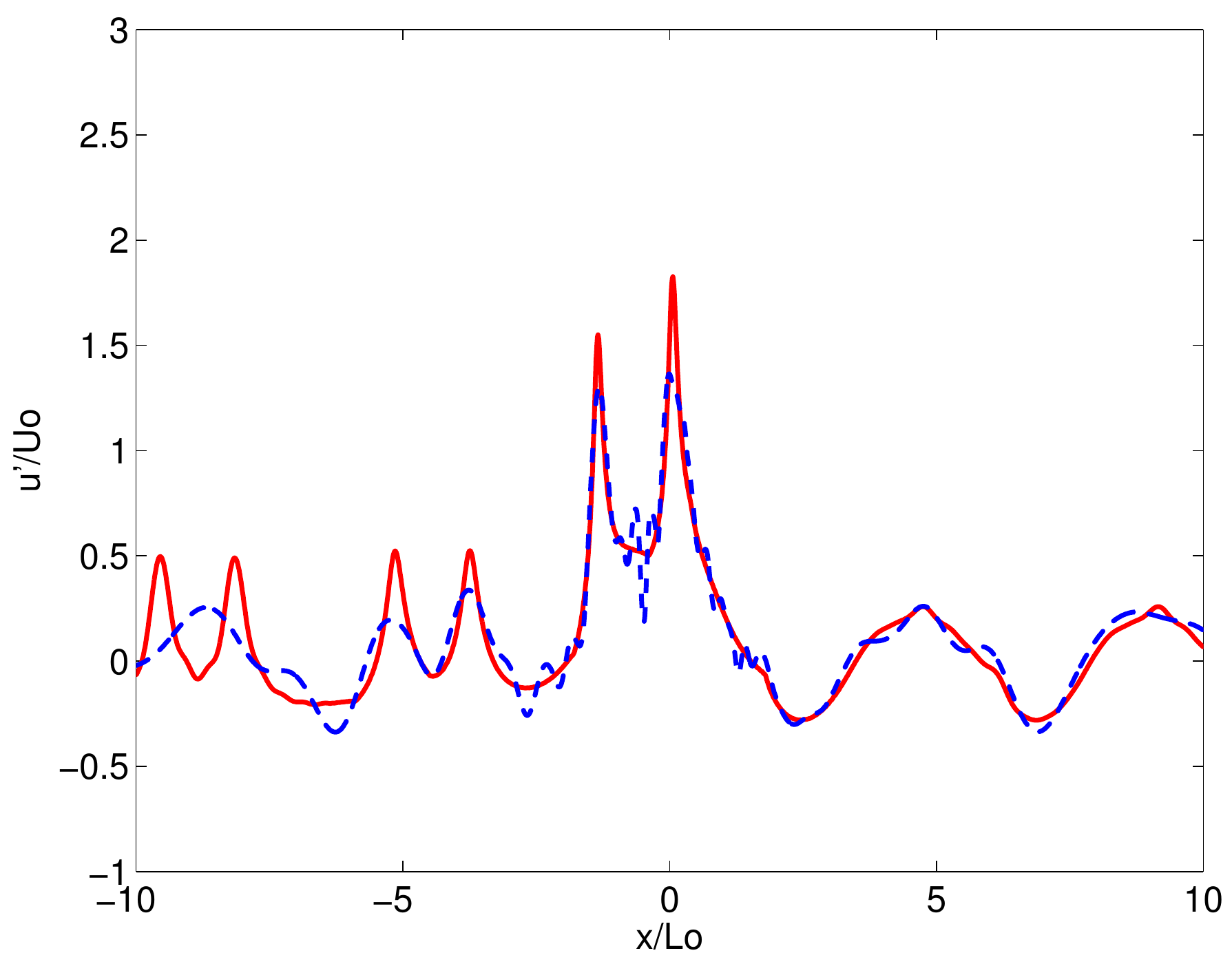}}
  \caption{Normalized horizontal velocity $u/U_0$ for tidal flow over isolated topography with uniform stratification: (a) vertical cross-section of nonlinear simulation results using NFA. The dashed line shows the beam path predicted by linear theory; (b) vertical cross-section of linear theory results; and (c) comparison of NFA with linear theory for a horizontal line at $z/L_0=-0.64$, where $L_0 = H - h_0$.}
\label{fig:UniformStratAsymHill}
\end{figure}

\begin{figure}[h!]
   \centerline{(a)\includegraphics[width=0.85\linewidth]{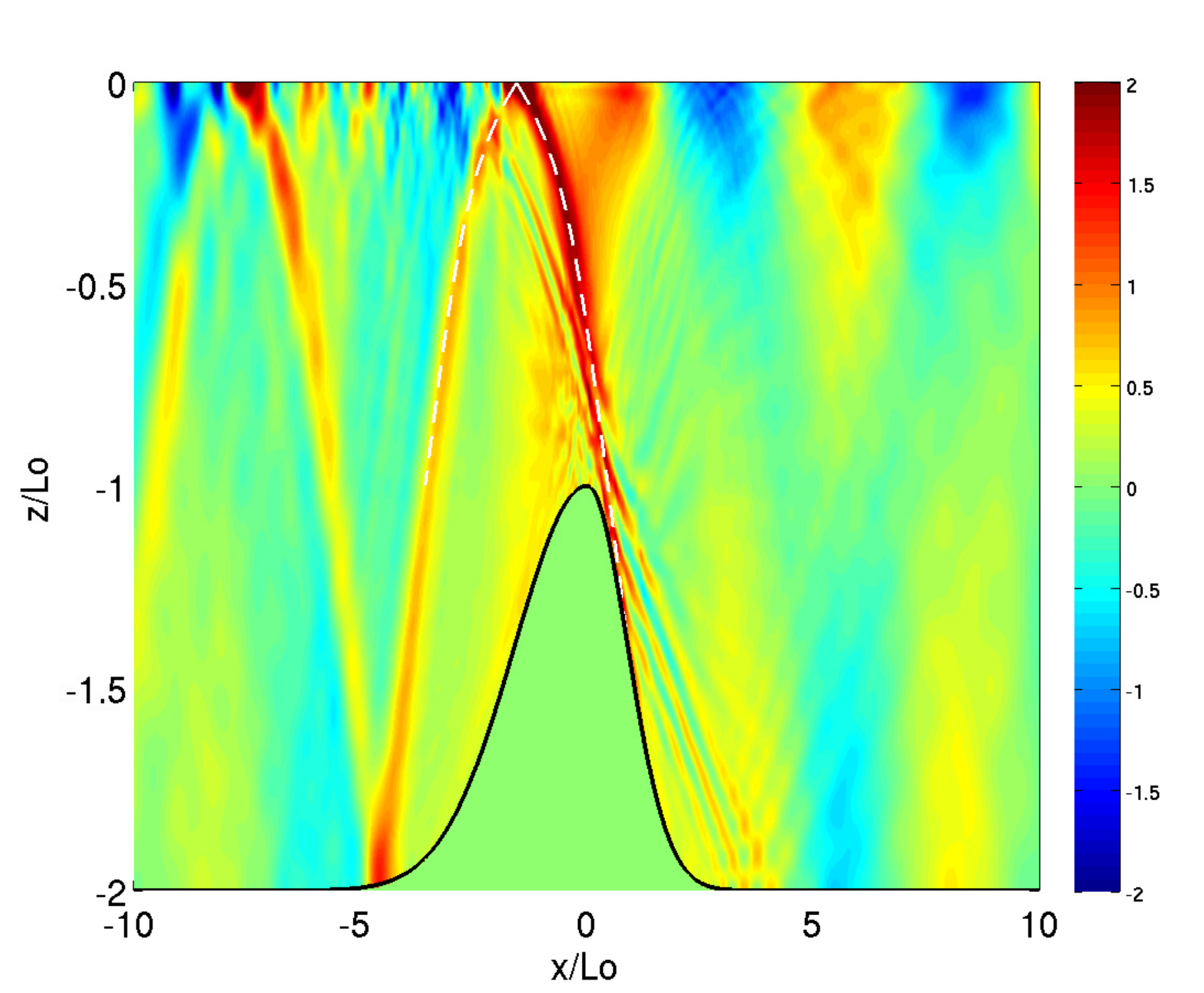}}
   \centerline{(b)\includegraphics[width=0.85\linewidth]{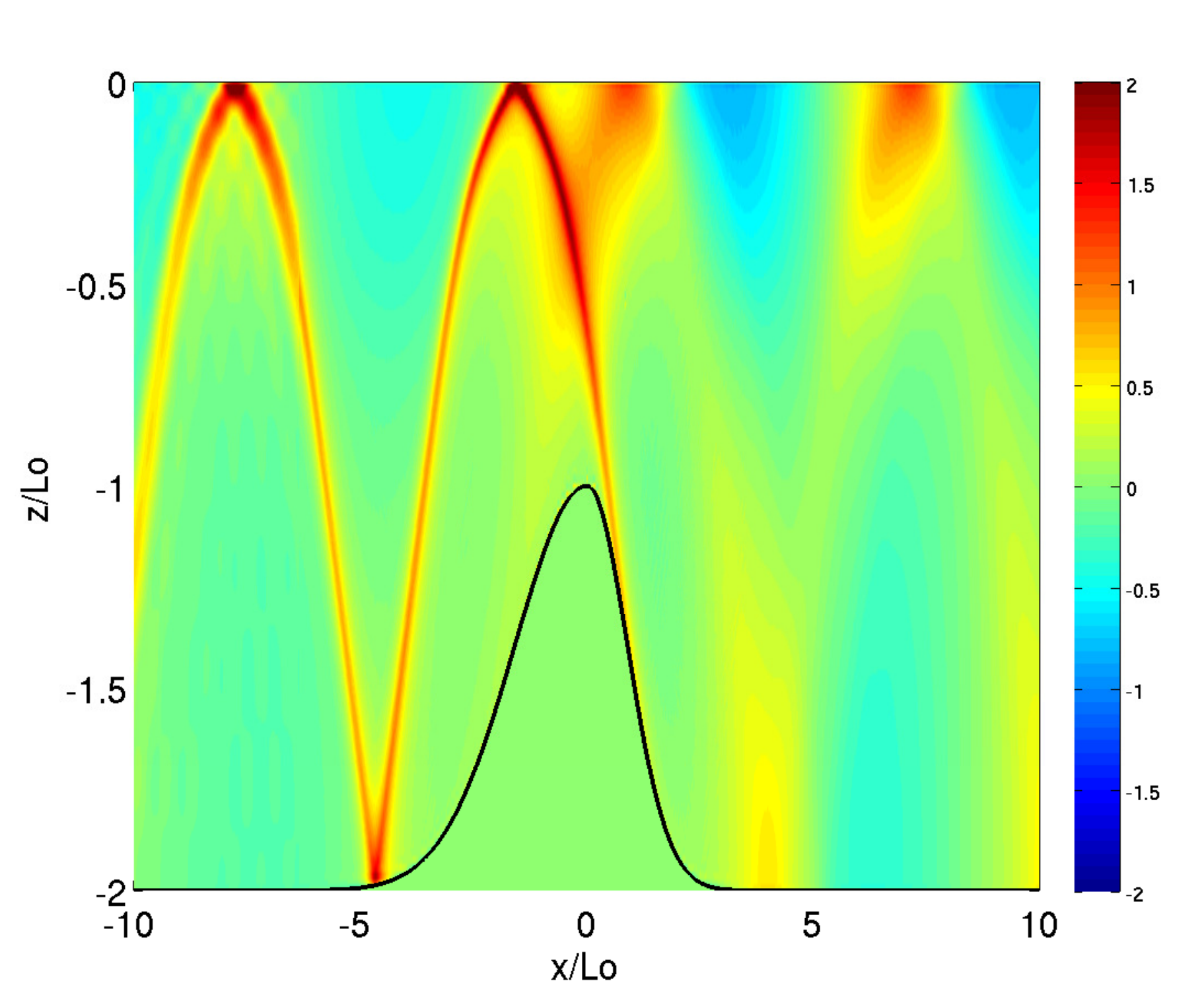}}
   \centerline{(c)\includegraphics[width=0.85\linewidth]{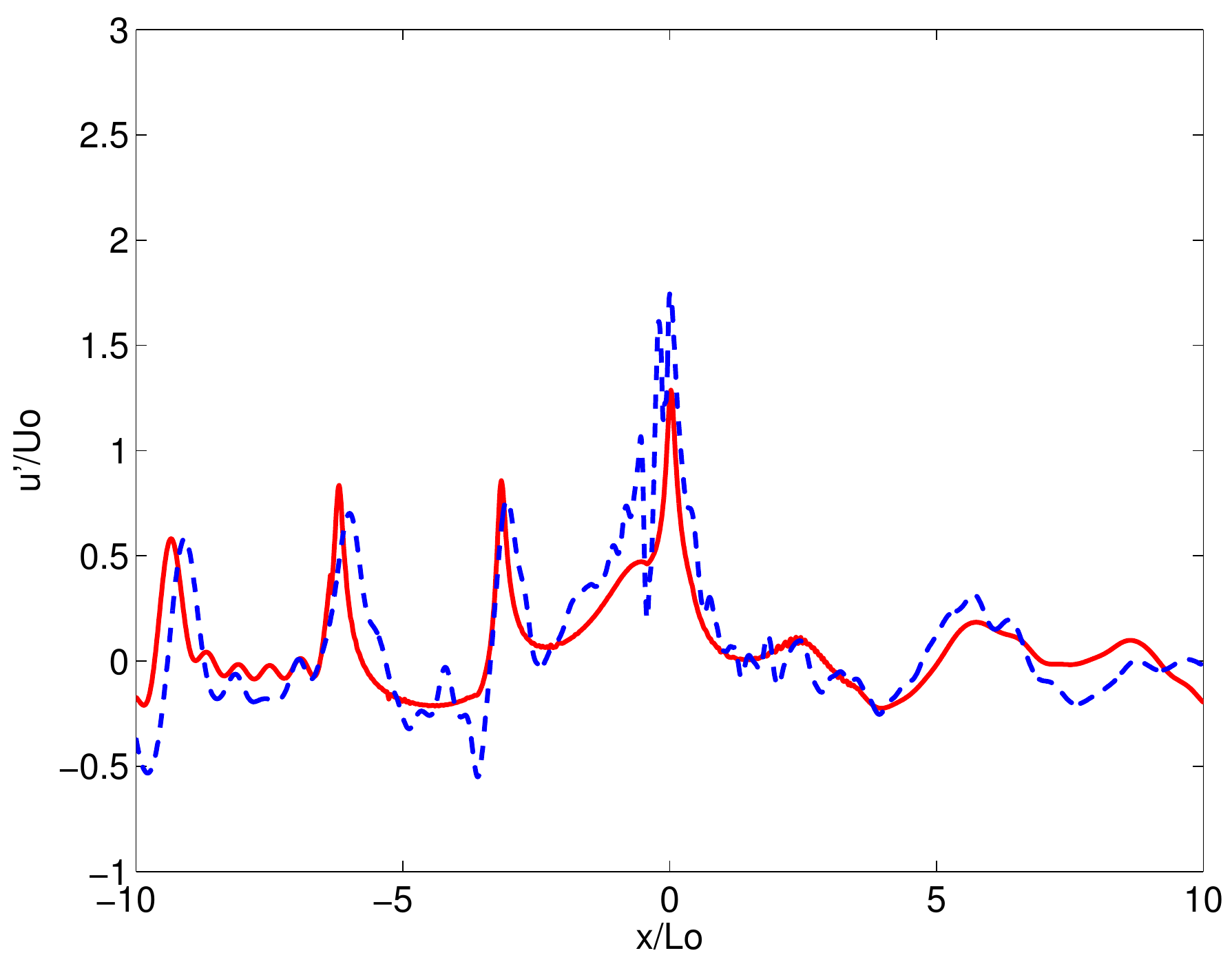}}
  \caption{Normalized horizontal velocity $u/U_0$ for tidal flow over isolated topography with $non$-uniform stratification: (a) vertical cross-section of nonlinear simulation results using NFA. The dashed line shows the beam path predicted by linear theory; (b) vertical cross-section of linear theory results; and (c) comparison of NFA with linear theory for a horizontal line at $z/L_0=-0.64$, where $L_0 = H - h_0$.}
\label{fig:NonUniformStratAsymHill}
\end{figure}

\subsection{Enforcement of No-Slip Boundary Conditions}

The stress $\tau_i$ is used to impose partial-slip and no-slip conditions on the surface of the bottom using a body-force formulation as follows.
\begin{equation}
\label{body}
\tau_i = \beta ( u_i -v_i) \delta ( {\bf x} - {\bf x_b^+} ) \; ,
\end{equation}
where $\beta$ is a body-force coefficient, $v_i$ is the velocity of the body, ${\bf x}$ is a point in the fluid, and ${\bf x_b^+}$ is a point slightly outside the body.  Equation~\eqref{body} forces the fluid velocity to match the velocity of the body.  Note that free-slip boundary conditions are recovered with  $\beta=0$ and no-slip boundary conditions are imposed as  $\beta \to \infty$.   

\citeasnoun{dommermuth98} discuss modeling using body-force formulations. Recently, there have been several studies which use similar body boundary conditions in both finite volume (\citeasnoun{Meyer:2010}, \citeasnoun{Meyer:2010a} and finite element (\citeasnoun{Hoffman:2006}, \citeasnoun{Hoffman:2006a}, \citeasnoun{John:2002} simulations.  The finite element simulations of \citeasnoun{Hoffman:2006a} at very high Reynolds numbers are able to predict the lift and drag to within a few percent of the consensus values from experiments, using a very economical number of grid points.  The finite volume implementations have also shown promise albeit at more modest Reynolds numbers ($Re=3,900$).  \citeasnoun{nfa3} show that finite volume simulations with $Re \to \infty$ are able to accurately predict flow about bluff-bodies using a partial-slip type boundary condition to model the
effects of the unresolved turbulent boundary layer.

\subsection{Formulation of Inflow and Outflow Boundary Conditions}

The effects of oscillating tidal currents can be modeled by including an oscillatory current in velocity:
\begin{equation}
\label{eq:current1}
u_1 \leftarrow u_1 + U^c \cos(\omega t).
\end{equation}
$U^c$ is the amplitude of the current at inflow normalized by $U_0$ and $\omega$ is the frequency of the current normalized by $L_0/U_0$.   

The velocities near the inflow and outflow boundaries are nudged every $N$ time steps toward target velocities that conserve flux:
\begin{eqnarray}
\label{eq:current2}
u_1 & \leftarrow & u_1 - \beta N \Delta t  f^I(x) (u_1 - U^I)   \; {\rm for} \;  x \in \; {\rm \vee_I}  \nonumber \\
u_1 & \leftarrow & u_1 - \beta N \Delta t  f^O(x) (u_1 - U^O)  \; {\rm for}  \;  x \in \; {\rm \vee_0}. \nonumber \\ & &
\end{eqnarray}
$\beta$ is a relaxation parameter, and $f^I(x)$ and $f^O(x)$ are tapering functions at the inflow and outflow regions, respectively.   $U^I$ and $U^O$ are the target velocities at the inflow and outflow boundaries, respectively.   $\vee_I$ and $\vee_0$ denote regions near the inflow and outflow boundaries, respectively.

Away from the inflow boundary,  $f^I(x)=0$.    At the inflow boundary, $f^I(x)=1$.   $f^I(x)$ smoothly transitions between the two regions.   Similarly, $f^O(x)=0$ away from the outflow boundary, and $f^O(x)=1$ at the outflow boundary.

At inflow, the target velocity is set to zero, $U^I=0$, then by conservation of flux, 
\begin{equation}
\label{eq:current3}
U^O = \left( \frac{S_I}{S_0} -1 \right) U^c \cos(\omega t)
\end{equation}
where $S_0$ is the surface area at the outflow boundary, and $S_I$ is the surface area at the inflow boundary.

\section{VALIDATION}

We desribe several preliminary simulations using NFA to assess the accuracy of the numerical scheme for simulating realistic internal wave tidal beams and their interactions with an ocean thermocline.

\subsection{Flow over a two-dimensional ridge: comparison with linear theory}

A first step to validate the numerical technique is to compare its results for an idealized two-dimensional problem with linear theory.  \citeasnoun{echeverri2010} developed a linear model using a Greens function approach to simulate internal tidal beams generated by tidal flow over two-dimensional topography with non-uniform stratification.  The relevant parameters in this case are $\omega$, the tidal frequency; $N$, the buoyancy frequency; $h_0$, the topographic height; $H$, the water depth; $L$, a horizontal length scale representative of the topography; and $U_0$, the barotropic tidal current amplitude. The governing nondimensional parameters are: the frequency ratio $\omega/N$; the relative height of the topography $h_0/H$; the tidal excursion parameter, defined as the tidal excursion distance normalized by the topographic length scale,  $\epsilon_{ex} = U_0/(\omega L)$; and the steepness parameter, defined as the ratio of the slope of the topography to the slope of the tidally generated internal waves,  $\epsilon_S = h_0/(L s)$, where s = $\sqrt{\omega^2/(N^2-\omega^2)}$.

We performed two-dimensional numerical simulations for tidal flow over an asymmetrical ridge with uniform stratification, as shown in figure \ref{fig:Density_hill}a, or nonuniform stratification, as shown in figure \ref{fig:Density_hill}b. The nonuniform stratification represents an idealized seasonal thermocline. The simulations have $\omega/N = 0.67$ and $h_0/H = 0.5$. For linear theory to be accurate, the flow must have a subcritical value for the steepness parameter ($\epsilon_S<1$) and a small value for the excursion parameter ($\epsilon_{ex}<<1$). In figures \ref{fig:UniformStratAsymHill} and \ref{fig:NonUniformStratAsymHill}, $\epsilon_S = 0.42, 0.75$ and $\epsilon_{ex} = 0.029, 0.062$, for the shallow slope side of the topography and steep slope side of the topography, respectively.  The steepness parameters are subcritical, but close enough to critical for some nonlinear effects to exist.

Figure  \ref{fig:UniformStratAsymHill}a shows results from an NFA simulation of uniformly stratified flow over a ridge after $10$ tidal periods of adjustment.  Visually, the comparison between NFA and linear theory (figure \ref{fig:UniformStratAsymHill}b) is very good. The tidal beam is generated at the steepest part of the slope and is seen as a straight line in the flow, since the background stratification is homogeneous. The dashed line in figure \ref{fig:UniformStratAsymHill}a shows the ray path predicted by linear theory, which agrees well with the primary beam shown in the nonlinear simulation. Figure \ref{fig:UniformStratAsymHill}c compares the nonlinear and linear results at $z/L_0=-0.64$, where $L_0 = H - h_0$.  While the nonlinear results show some nonlinearly generated harmonics, which of course do not exist in the linear theory, the magnitudes of the beams and the large-scale behavior compare very well.

Figure \ref{fig:NonUniformStratAsymHill} shows a similar computation for $non$-uniform stratification.  This stratification is more realistic, with an idealized thermocline profile near the surface. Figure \ref{fig:NonUniformStratAsymHill}a shows results from NFA after $18$ tidal periods of adjustment.  When comparing it with linear theory (figure \ref{fig:NonUniformStratAsymHill}b), the NFA results have a slightly stronger magnitude.  We believe this stronger amplitude is due to nonlinearities in the flow caused by the strong stratification near the surface.  The dashed line in figure \ref{fig:NonUniformStratAsymHill}a shows the ray path predicted by linear theory, which agrees well with the primary beam shown in the nonlinear simulation.  Figure \ref{fig:NonUniformStratAsymHill}c compares the nonlinear and linear results at $z/L_0=-0.64$.  Linear theory captures the main flow behavior, but the nonlinear simulation shows some higher frequency perturbations about the main flow.

The process of energy transfer from semi-diurnal internal tides (M2) to subharmonic internal tides (M1) and higher harmonics (M4, M6, etc.) has been a topic of recent interest.  \citeasnoun{gerkema2006} discusses how energy can transfer to subharmonic internal tides which produce features that appear as ``slices'' in the generated internal tidal beam.  As time goes on, more and more ``slices'' appear.  The slices are the troughs and crests of the M1 beams.  Higher harmonics are generated when the beam reflects off the bottom due to the nonlinear interaction of the incoming and outgoing beams.  We will show that this is also true for beams that reflect off the surface and thermocline.

\begin{figure}[h!]
     \centerline{(a)\includegraphics[width=0.85\linewidth]{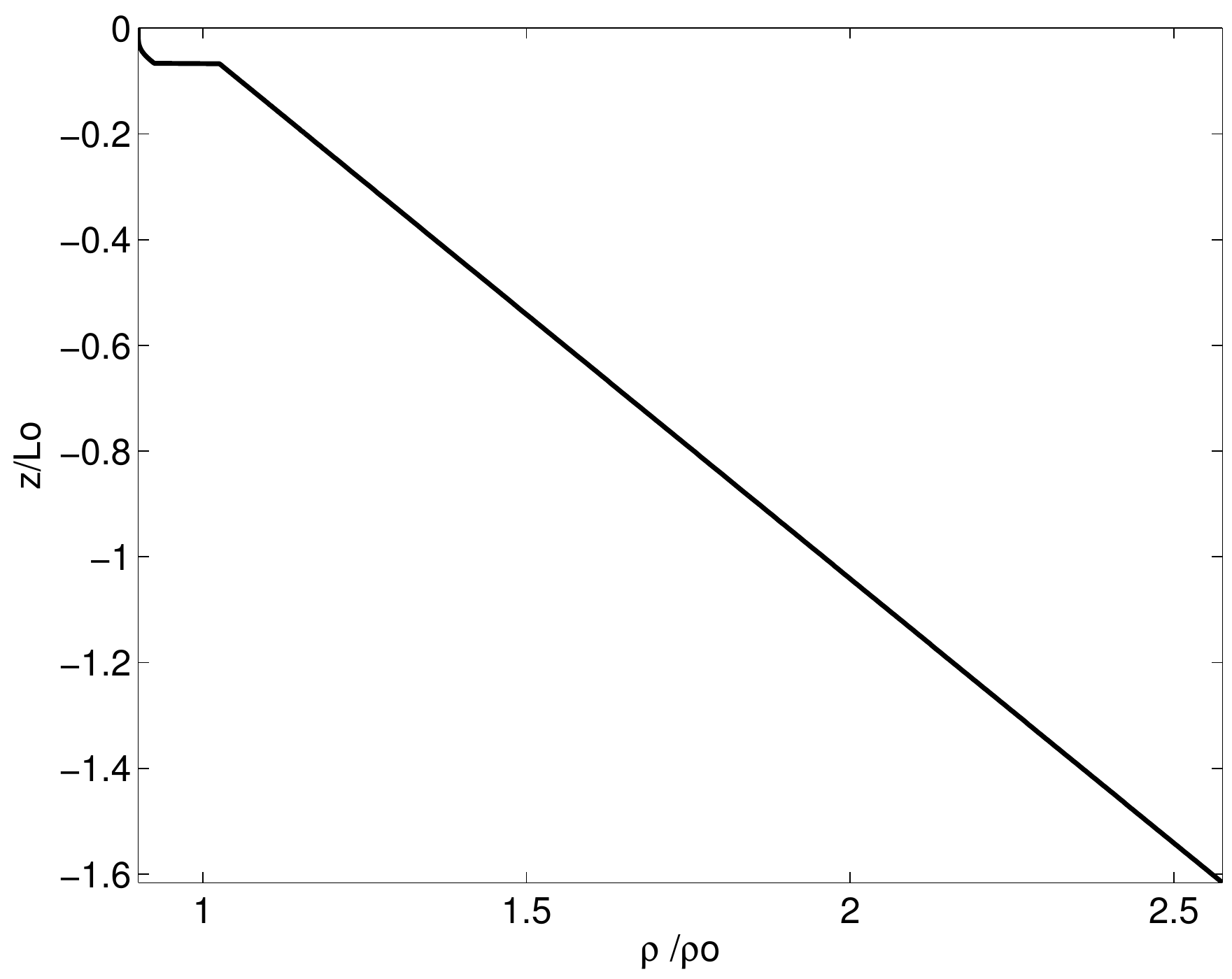}}
  \caption{Normalized density $\rho/\rho_0$ as a function of normalized depth $z/L_0$, where $L_0 = H - h_0$: nonuniform stratification for a thermocline (approximated as a density jump) over a continental shelf.}
\label{fig:Density_shelf}
\end{figure}

\begin{figure*}[h!]
  \centerline{
     \includegraphics[width=0.5\linewidth]{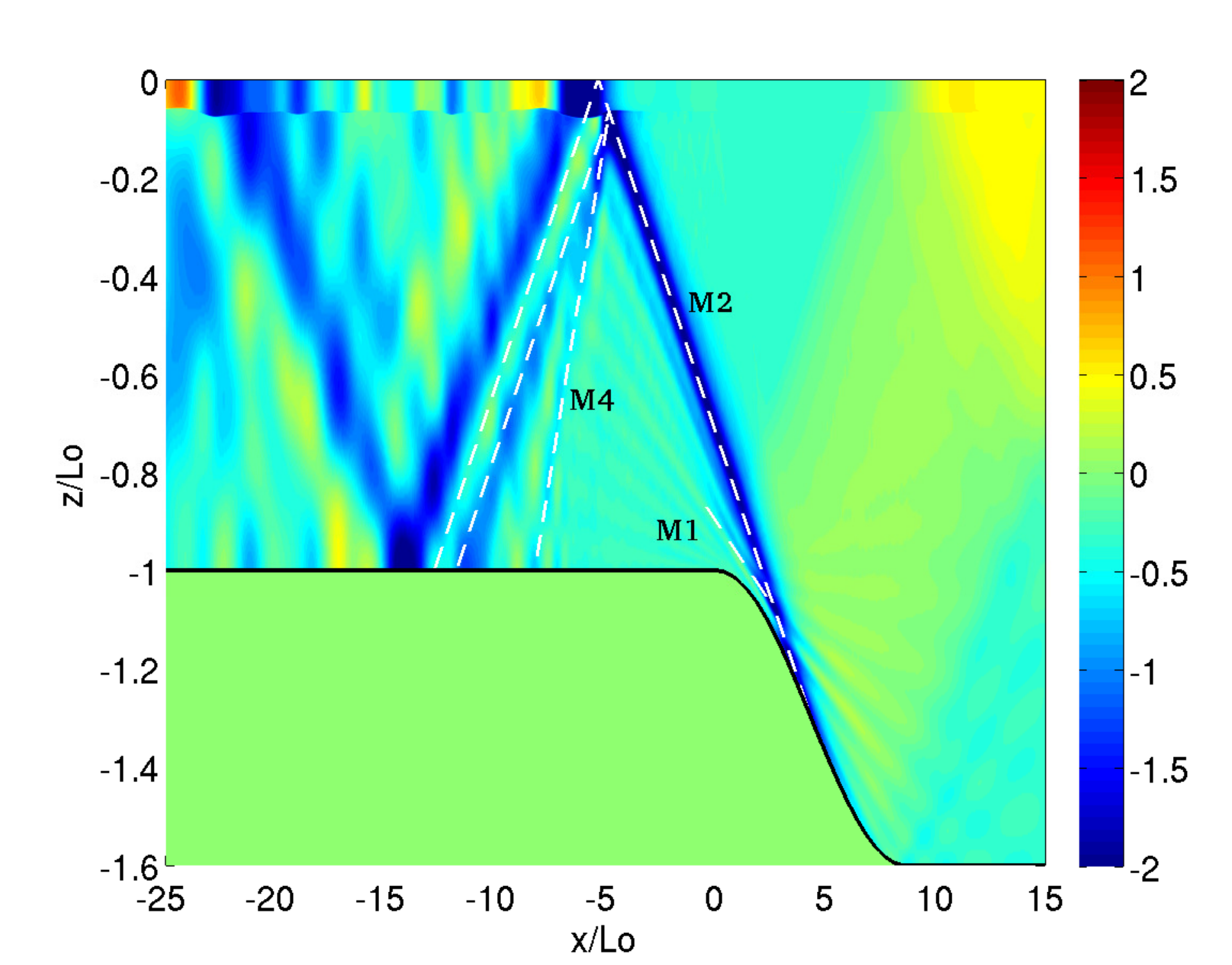}
     \includegraphics[width=0.5\linewidth]{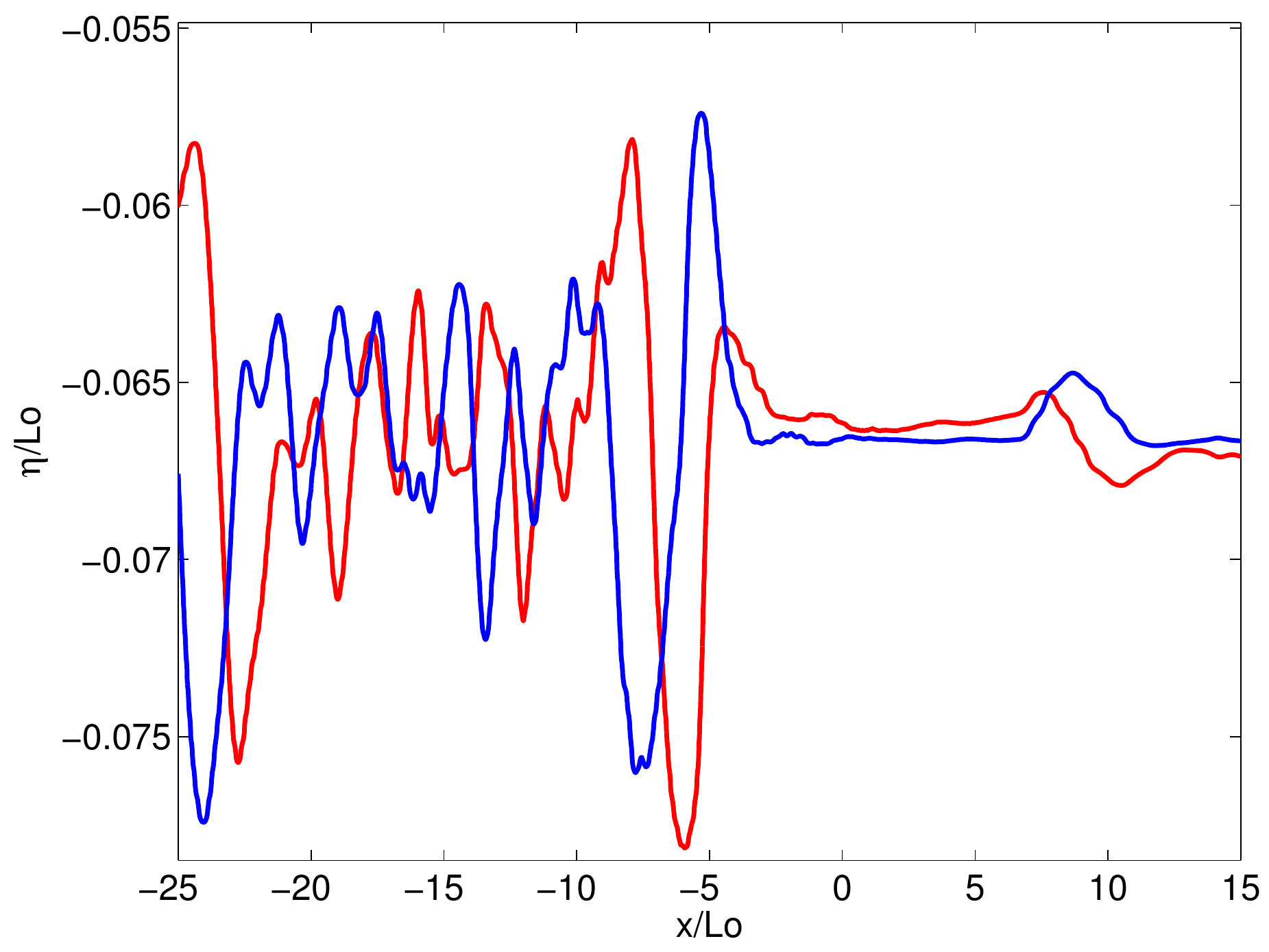}
   }
  \centerline{
  {\small (a)}\hspace{2.5in}{\small (b)}
  }
  \caption{(a) Contours of $u/U_0$ after $8.25$ tidal periods, depicting the generation of an internal wave tidal beam and its interaction with the thermocline.  The beam paths associated with the M1, M2 and M4 tidal modes are shown as white dashed lines. (b) Elevation of thermocline: (red) at $8.25$ tidal periods (a); (blue) at $8.5$ tidal periods}
\label{fig:IWTB_ISW}
\end{figure*}

\subsection{Flow over a two-dimensional continental shelf}

NFA results for the internal wave tidal beam generated by the interaction of the barotropic tide with an idealized, two-dimensional continental shelf break, with the vertical density profile shown in figure \ref{fig:Density_shelf}, is shown in figure \ref{fig:IWTB_ISW}a.  In this example, $\omega/N = 0.13$; $h_0/H = 0.375$;  the tidal excursion parameter is $\epsilon_{ex} = 0.0046$; and the peak steepness parameter is $\epsilon_S = 0.81$.  Note that the tidal frequency in this example is two orders of magnitude higher than it is in the real ocean. This was done as a computationaly inexpensive way to obtain a regime in which internal solitary waves are formed by internal wave tidal beams (for testing purposes). In future studies, other flow parameters will be adjusted to reach this regime with the tidal frequency set to the semi-diurnal tidal frequency.  

The NFA numerical domain was 8192x512 grid points with $\Delta x /L_0 = 0.007$ and the minimum $\Delta z / L_0 = 0.0007$  and maximum $\Delta z / L_0 = 0.007$.  There were approximately $188$ grid points across the interface.  There were $170$ and $703$ grid points across the bathymetry in the $z$-direction and $x$-direction, respectively.

The tidal beams associated  with M1, M2 and M4 harmonics, according to linear theory, are plotted as white dashed lines. Close to the generation site,  parametric subharmonic instability causes the observed fan of weaker beams about the main beam, while further away where the beam reflects from the thermocline, or the lower boundary, higher harmonics are generated due to the nonlinear interaction of incoming and reflected beams.  These instabilities grow over time and can eventually cause the internal wave tidal beam to break.

The interaction of the internal wave tidal beam with the thermocline also generates internal solitary waves (ISW) that propagate horizontally along the thermocline.   Figure \ref{fig:IWTB_ISW}b shows the vertical displacement of the thermocline as a function of $x$ at two successive times. Comparing the red line with the the blue line in figure \ref{fig:IWTB_ISW}b at $x/L_0\approx -13$, a lead wave and successive smaller waves are seen propagating up the shelf.  A rough estimate of the wave velocity is $C/C_0 = 1.123$, where $C_0$ is the linear long-wave speed for internal waves in a two-layer fluid, which is consistent for a two-layer solitary wave with the shown amplitude.

\section{CONCLUSIONS} \label{sec:conclusions}

We have described the formulation and implementation of a fully nonlinear, three-dimensional numerical model that is capable of simulating with great accuracy tidal flow of a stratified ocean over arbitrary bottom topography, even though the range of scales from the long scales of the shelf topography to the fine scales of the resonantly generated internal wave tidal beams is very large. This new numerical model is an extension of the existing NFA code, which originally was designed for the computation of the motion of ships in a free-surface flow. Furthermore, we have used the newly modified numerical model to compute tidal flow over an idealized two-dimensional ridge and found the simulations to agree very well with linear theory for both uniform and nonlinear stratification. Finally, the model was used to generate a tidal beam by the interaction of the barotropic tide with an idealized two-dimensional continental shelf break. The simulation results show the generated beam and its harmonics, generated by nonlinear interactions, as well as the solitary waves generated by the beam's interactions with the thermocline.
\section{Acknowledgements}
This research was sponsored by Dr. Ron Joslin at the Office of Naval Research (contract number N00014-08-C-0508), Dr. Tom C. Fu at the Naval Surface Warfare Center, Carderock Division, and SAIC's Research and
Development program.  The numerical simulations were supported in part by a grant from the Department of Defense High Performance Computing Modernization Program (\mbox{http://www.hpcmo.hpc.mil/}).  The numerical simulations were performed on the SGI\textsuperscript{\textregistered} Altix\textsuperscript{\textregistered} ICE 8200LX (Silicon Graphics International Corporation) at the U.S. Army Engineering Research and Development Center. Animated versions of the numerical simulations described here as well as many others are available online at \url{http://www.youtube.com/waveanimations}

%
%
\bibliographystyle{29onr}
\bibliography{29onr}
\end{document}